\documentclass[a4paper,aps,prl,reprint,superscriptaddress, twocolumn,preprintnumbers,amsmath,amssymb,nobalancelastpage]{revtex4-1}
\usepackage{amsmath}
\usepackage{verbatim}
\usepackage{graphicx}
\usepackage{color}
\usepackage[normalem]{ulem}
\usepackage{amsmath,amsfonts,amssymb,graphicx,color,times,bbm,psfrag,dsfont,slashed,bm}
\usepackage{gensymb}
\usepackage{natbib}
\usepackage[T1]{fontenc}
\usepackage[latin9]{inputenc}

\newcommand{\beginsupplement}{%
        \setcounter{table}{0}
        \renewcommand{\thetable}{S\arabic{table}}%
        \setcounter{figure}{0}
        \renewcommand{\thefigure}{S\arabic{figure}}%
     }

\begin{document}

\title{Simulating twistronics without a twist}

\author{Tymoteusz Salamon}
\affiliation{ICFO - Institut de Ciencies Fotoniques, The Barcelona Institute of Science and Technology,
Av. Carl Friedrich Gauss 3, 08860 Castelldefels (Barcelona), Spain}
\author{Alessio Celi}
\affiliation{Departament de F{\'i}sica, Universitat Aut{\`o}noma de Barcelona, 08193 Bellaterra, Spain}
\author{Ravindra W. Chhajlany}
\affiliation{Faculty of Physics, Adam Mickiewicz University, 61614 Poznan, Poland}
\author{Ir{\'e}n{\'e}e Fr{\'e}rot}
\affiliation{ICFO - Institut de Ciencies Fotoniques, The Barcelona Institute of Science and Technology,
Av. Carl Friedrich Gauss 3, 08860 Castelldefels (Barcelona), Spain}
\affiliation{Max-Planck-Institut f{\"u}r Quantenoptik, D-85748 Garching, Germany}
\author{Maciej Lewenstein}
\affiliation{ICFO - Institut de Ciencies Fotoniques, The Barcelona Institute of Science and Technology,
Av. Carl Friedrich Gauss 3, 08860 Castelldefels (Barcelona), Spain}
\affiliation{ICREA, Pg. Lluis Companys 23, Barcelona, Spain}
\author{Leticia Tarruell}
\affiliation{ICFO - Institut de Ciencies Fotoniques, The Barcelona Institute of Science and Technology,
Av. Carl Friedrich Gauss 3, 08860 Castelldefels (Barcelona), Spain}
\author{Debraj Rakshit}
\affiliation{ICFO - Institut de Ciencies Fotoniques, The Barcelona Institute of Science and Technology,
Av. Carl Friedrich Gauss 3, 08860 Castelldefels (Barcelona), Spain}
\affiliation{Max-Planck-Institut f{\"u}r Quantenoptik, D-85748 Garching, Germany}

\date{\today}
\begin{abstract}

Rotational misalignment or twisting of two mono-layers of graphene  strongly influences its electronic properties. Structurally, twisting leads to large periodic supercell structures, which in turn can support intriguing strongly correlated behaviour. Here, we propose a highly tunable scheme to synthetically emulate  twisted bilayer systems with   ultracold atoms trapped in an optical lattice. In our scheme,  neither a  physical bilayer nor  twist is directly realized. Instead, two synthetic layers are produced exploiting coherently-coupled internal atomic states, and a supercell structure is generated \emph{via} a spatially-dependent Raman coupling. 
To illustrate this concept, we focus on a synthetic square bilayer lattice and show
that it leads to tunable quasi-flatbands and Dirac cone spectra under certain magic supercell periodicities. 
The appearance of these features are explained using a perturbative analysis. 
Our proposal can be implemented using available state-of-the-art experimental techniques, and opens the route towards the controlled study of strongly-correlated flat band accompanied by hybridization physics akin to magic angle bilayer graphene in cold atom quantum simulators.
\end{abstract}

\maketitle

Novel routes to band engineering in lattice systems often lead to fundamental questions and new material functionalities. Different schemes of stacking 2--dimensional layers have emerged as a fruitful way of modifying material properties through the design of supercell structures and opened the field of so-called Van der Waals materials~\cite{Geim2013}. In particular, twisted bilayer graphene (TwBLG) has emerged as a tunable experimental platform hosting flat band physics and  strongly-correlated phenomena, such as possibly unconventional superconductivity, magnetism and other exotic phases \cite{Cao18-1, Cao18-2, Yankowitz19,Lu19}. This has inspired much theoretical debate around the origin of the electronic properties of TwBLG \cite{Volovik18,Yuan18,Koshino,Ochi,Zou18,Peltonen18,Padhi18,Sboychakov19,Guinea18,Xu18,Wu18,Isobe18,You18,Lian19,Lin19,Song19}.

The interesting correlated phenomenology is apparently related to Moir{\'e} patterns around small twist angles,  the so-called \emph{magic angles}, which lead to band flattening or effective mass reduction already at the single-particle level \cite{Morell10,Bistritzer11,Santos12,Koshino12,Tarnopolsky19}.  The geometrical Moir{\'e} patterns  physically induce spatially varying interlayer couplings that are behind the strong modification of the band structure. As in artificial graphene systems \cite{Polini2013}, emulating this physics beyond materials research may allow identifying key minimal ingredients that give rise to the phenomenology of TwBLG while also providing additional microscopic control. Photonic systems are particularly suited to explore this physics at the single-particle level. Very recently, single-particle transport in tunable photonic Moir\'e lattices has been experimentally studied \cite{Wang19}, where two dimensional localization of light and localization-delocalization have been experimentally demonstrated. Ultracold atoms in optical lattices  \cite{Bloch08,Lewenstein12} are the most promising platform to explore expeirmentally also the corresponding emerging many-body phenomena. The experimental realization of artificial graphene geometries \cite{Soltan-Panahi11,Tarruell12}, lattice geometries displaying flat bands like Kagome \cite{Jo12} and Lieb \cite{Taie15,Ozawa17}, or quasi-crystal structures \cite{Guidoni97,Viebahn18,Rajagopal19}, provides the building blocks for such exploration.

One obvious approach to study twisted bilayer graphene physics with ultracold atoms is to directly implement twisted bilayers using two intertwined optical lattices, as recently proposed in Ref. \cite{Tudela19}. Schemes for simulating other bilayer heterostructures have also been put forward \cite{Grass16}. This direct strategy poses significant experimental challenges, as it is difficult to stabilize the two layers at relative small angles and simultaneously achieve a sufficiently large lattice containing several supercells. Here, we propose an  alternative  scheme that builds on the concept of synthetic dimensions, i.e.,  reinterpreting the coherent Raman coupling between spin states of an atom as controllable tunneling along an artificial extra dimension \cite{Boada12, Celi13, Ozawa19}. 
For this, we note that the key effect of physical twisting is to induce  spatially modulated interlayer tunnelings across the lattice. In our scheme,  the  modulated interlayer  tunneling patterns are proposed to be directly imprinted  on the lattice \emph{via} spatial control of  the Raman couplings,   amounting thus to twisting the system without a physical twist. 
This leads to the creation of supercells with controllable sizes and shapes. By adjusting the strength, phase and spatial periodicity of $\Omega$, we show that an exemplary   bilayer square lattice system  supports a broad range of band structures. In particular, \emph{magic} values of the periodicity result in the appearance of (quasi-) flat bands as well as Dirac cone spectra. Although we focus in our illustrative example on a particular spatial modulation, general  interlayer coupling patterns can be experimentally induced including quasi-perioidic or Moir{\'e}-like patterns. 
Our proposal can be realized with fermionic two-electron atoms, such as strontium or ytterbium, using available experimental techniques.  \\

\begin{figure}[t!]
\centering
\includegraphics[clip,width=0.89\columnwidth]{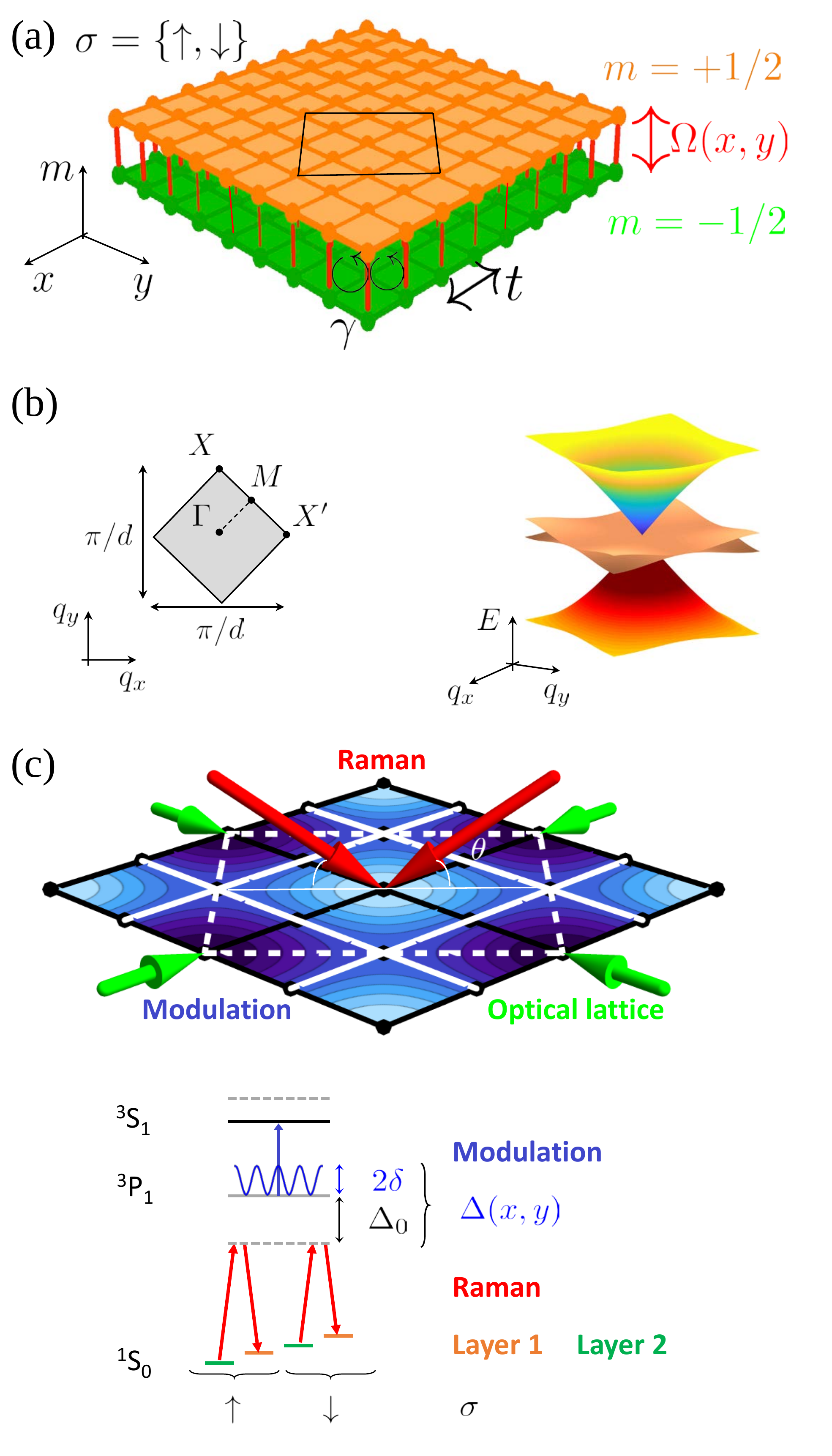}
\caption{{ Synthetic bilayer structure with a  supercell.} (a) Real-space potential of the synthetic bilayer. Each plane corresponds to one spin state $m=\pm 1/2$ (orange, green), which experiences a square lattice potential (tunneling $t$) and is connected to the other layer by a spatially-dependent and complex coupling $\Omega(x,y)$ (vertical red lines of variable width).
A top view of the lattice indicating the unit cell of the system containing $2\times 8$ sites for $l_x=l_y=4d$ is shown (black line). (b) Sketch of the first Brillouin zone, indicating the position of the high-symmetry points, and three dimensional view of the energy spectrum in the vicinity of $E=-\Omega_0(1-\alpha)$ for $\Omega_0 \alpha/h=20 t$ with $\alpha=0.2$ and $\gamma=0$. It has two quasi-flat bands intersecting a Dirac point. Note that a simple square lattice supports neither flat-bands nor Dirac cones. (c) Proposed experimental realization. Top: Two retroreflected optical lattice beams (green) create the square lattice. Two Raman beams of opening angle $\theta$ (red) produce  complex synthetic tunneling between the two layers. One ``modulation laser'' with a spatially varying intensity distribution (blue) modulates the amplitude of the Raman coupling. Bottom: laser beams involved in the synthetic bilayer coupling scheme. The single-photon detuning of the Raman beams (red arrows) is spatially modulated with respect to its initial value $\Delta_0/2\pi\sim 75$ MHz using a laser beam blue detuned with respect to the $^3P_1\rightarrow\, ^3S_1$ transition (blue arrow). It produces a light shift of maximal amplitude $2\delta/2\pi\sim30$ MHz.}
\label{Fig1}
\end{figure}

\vspace{-0.4cm}
{\it Concept.} We consider a two-dimensional Fermi gas with four internal states, indicated here generically as spin states $\{m,\sigma\}=\pm 1/2$. The system is loaded into a spin-independent square optical lattice of lattice spacing $d$, which lies in the $x-y$ plane and is characterized by a real tunneling amplitude $t$. We select two states to play the role of the electron spin $\sigma=\,\,\uparrow$, and the other two of electron spin $\sigma=\,\,\downarrow$. In addition, spin states corresponding to the same $\sigma$ are coupled in pairs. We label them by the index $m$, and make them play the role of a synthetic layer dimension. Since $m=\pm 1/2$, we obtain a bilayer structure of synthetic layer tunneling given by the coherent coupling. In order to obtain a lattice geometry with a tunable supercell we choose the amplitude of the synthetic layer tunneling  to be spatially modulated according to $\Omega(x,y) = \Omega_0 \left[1 -\alpha(1+\cos{(2 \pi x/l_x)} \cos{(2 \pi y/l_y)}) \right]$. Here $l_x$ ($l_y$) is its periodicity along the $x$ ($y$) axis. The synthetic tunneling also induces a Peierls phase $\bm{\gamma}\cdot \bf r$, where $\bm{\gamma}=\gamma(\hat{x}+\hat{y})$ and $\bm{r}=x\hat{x}+y\hat{y}$. This mimics the effect of a magnetic flux that pierces the system perpendicularly to the synthetic layer dimension \cite{Celi13}. As depicted in Fig. \ref{Fig1}(a), the complete scheme represents a synthetic spinfull bilayer structure subjected to a magnetic field, denoted as $\Theta(l_x,l_y)$.

The Hamiltonian of the system is given by
\begin{align}
\label{eq:H}
H &= H_\mathrm{in}+  H_\mathrm{inter} \nonumber\\
&= -t \sum_{{\bf r},m,\sigma} \left[a_{m,\sigma}^{\dagger}({\bf r}+d\,\hat{x}) + a_{m,\sigma}^{\dagger}({\bf r}+d\,\hat{y}) \right.\\
& \hspace{1.1cm} \left. + \,\Omega({\bf r}) \exp(-i { \bm \gamma}\cdot {\bf r})~a_{m+1,\sigma}^{\dagger}({\bf r}) \right] a_{m,\sigma}({\bf r}) +\mathrm{H.c.}, \nonumber
\end{align}
where we distinguish the in-layer and the inter-layer tunnelings.

To diagonalize it, we combine a gauge transformation and a Fourier transform such that $a_{m,\sigma}({\bf r})=\sum_{\bf q} \exp\left( i({\bf q}\cdot{\bf r}+m{\bm \gamma}\cdot{\bf r})\right) a_{m,\sigma}({\bf q})$. Here, ${\bf q}$ is the momentum conjugated to ${\bf r}$. The Hamiltonian
can then be rewritten as $H=\sum_{{\bf q}} H_{{\bf q}}$, where the dimension of $H_{\bf q}$ is set by the spatial periodicity of the synthetic tunneling. Fig.~\ref{Fig1}(b) sketches the Brillouin zone of the bilayer system and a three-dimensional view of its energy spectrum for $l_x=l_y=4 d$, corresponding to $\Theta(4,4)$, for $\gamma = 0$. In the vicinity of $E=\pm \Omega_0(1-\alpha)$, it features two quasi-flat bands and a Dirac point touching them (only one of them is represented in Fig.~\ref{Fig1}(b)). This band structure is reminiscent of that of magic angle twisted bilayer graphene.

{\it Experimental proposal.} For specificity, we focus on the realization of this scheme employing a subset of four states out of the large nuclear spin manifold $I=9/2$ of $^{87}$Sr. Note however that our proposal is directly transposable to $^{173}$Yb ($I=5/2$). Thanks to the $SU(N)$ invariant interactions characteristic of two-electron systems, collisional redistribution of the atoms among the different states is inhibited. We select two of them to play the role of the electron spin $\sigma=\,\,\uparrow$, and the other two of spin $\sigma=\,\,\downarrow$. All are subjected to a two-dimensional spin-independent optical lattice potential, created by two counter-propagating lattice beams. We choose $\lambda_L=813$ nm, which is commonly used because it corresponds to the magic wavelength of the clock transition $^1S_0\rightarrow\,^3P_0$. We set a lattice depth $8\,E_L$, which yields $t/h=107$ Hz. Here, $E_L=\hbar^2 k_L^2/2m$ is the lattice recoil energy, $k_L=2\pi/\lambda_L$, and $d=\lambda_L/2$.

To create the synthetic layer tunneling, we exploit two-photon Raman transitions between spins $m=\pm 1/2$ \cite{footnote1}. We employ a pair of Raman beams of wavelength $\lambda_R=689$ nm near-resonant to the intercombination transition  $^{1}S_0\rightarrow\, ^{3}P_1$, which produce a coupling of amplitude $\Omega_0=\Omega_1\Omega_2/\Delta_0$. Here $\Omega_1$ and $\Omega_2$ are the individual coupling amplitudes of the Raman lasers and $\Delta_0$ the single-photon detuning. The Raman beams propagate in a plane perpendicular to the lattice potential, are aligned along its diagonal, and form an angle $\theta$ with the lattice plane (see Fig. \ref{Fig1}(c)). This yields an in-plane momentum transfer per beam $k_R=\pm2\pi \cos{\theta}/\lambda_R$, with projections $k_R/\sqrt{2}$ along the lattice axes. Therefore, the phase of the synthetic tunneling is $\bm{\gamma}\cdot{\bf r}=\gamma (x \hat{x}+ y \hat{y})$, with $\gamma=\pm2\pi \cos{\theta} \lambda_L/(\sqrt{2}\lambda_R)$. The sign is determined by the relative detuning of the Raman lasers. Experimentally, the simplest choice is to use counterpropagating Raman beams ($\theta=0\degree$), which yields $\gamma=0.8$ (mod $2\pi$). However, other magnetic fluxes can be easily realized by adjusting the value of $\theta$.
\begin{figure}[t]
\centering
\includegraphics[angle=0,width=6.7cm,height=8cm]{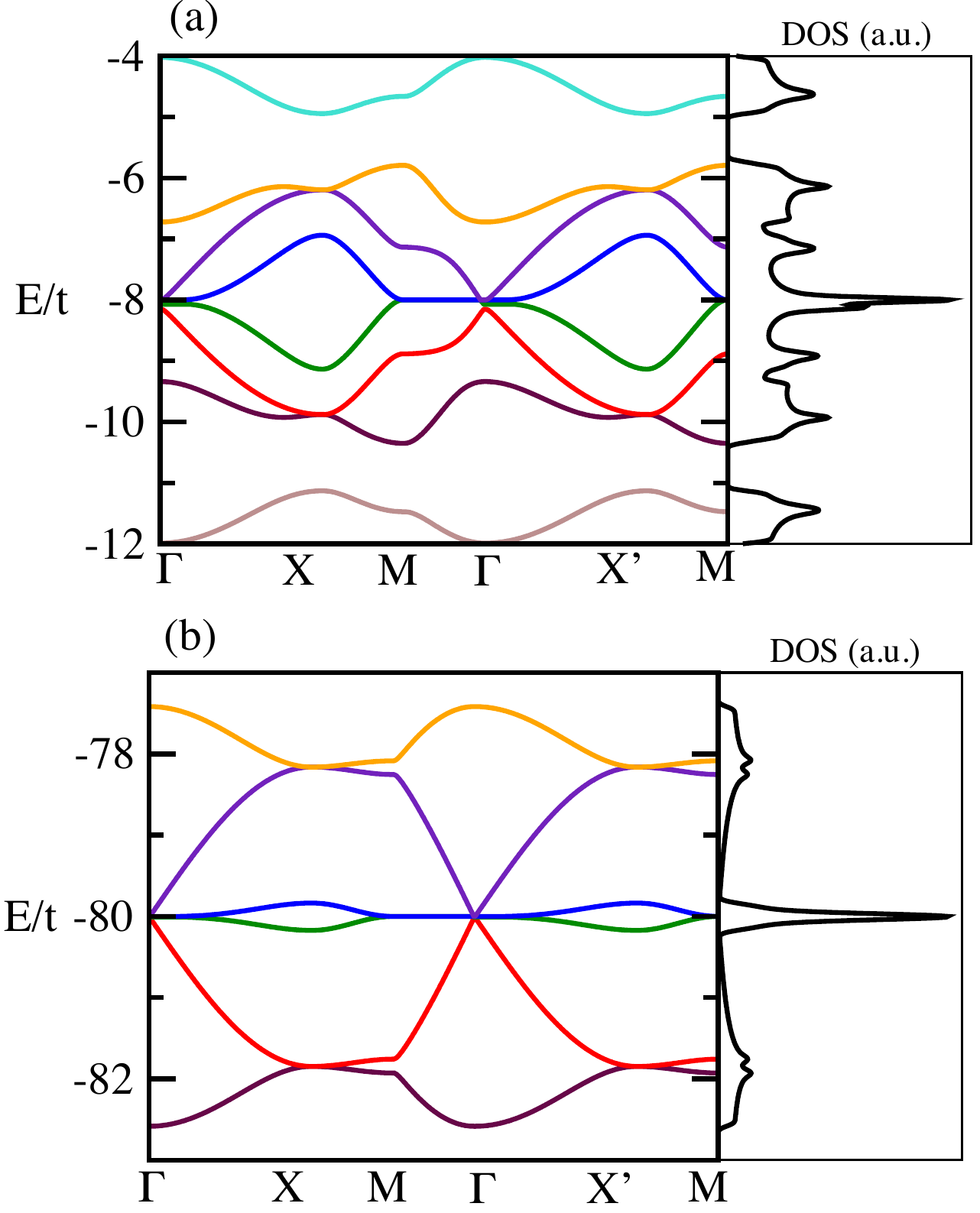}
\caption{{Magic configuration band structure and DOS.} Band structures around energy $-\Omega_0 (1-\alpha)$ and DOS (in arbitrary units) corresponding to $\Theta(4,4)$ supercell along the paths passing through the high-symmetry points ${\bf \Gamma, X, M, \Gamma, X', M}$. Panels (a) and (b) corresponds to $\Omega_0 \alpha/h=2 t$ and $20 t$, respectively, with $\alpha=0.2$ and $\gamma=0.8$. In the evolution from panel (a) to (b) the six central bands in panel (a), denoted with colors from orange to maroon, remain close in energy (as part of one single band, see perturbative analysis in \cite{eff-ni}) while the remaining two bands in panel (a) (in cyan and brown) separate in energy and do not appear in panel (b).}
\label{Fig2}
\end{figure}

To implement a periodic modulation of the Raman coupling amplitude on the scale of several lattice sites, which is the key ingredient of our scheme, we propose to exploit a periodic potential created by a laser close-detuned from the excited state to excited state transition $^{3}P_1\rightarrow\,^{3}S_1$ (corresponding to $688$ nm \cite{Stellmer13}). This results in a large light shift of the $^{3}P_1$ excited state of amplitude $\delta$, leading to a detuning of the Raman beams $\Delta (x,y)=\Delta_0+\delta (1+\cos{(2\pi x/l_x)} \cos{(2\pi y/l_y)})$. Its effect is to modulate the Raman coupling amplitude $\Omega(x,y)\simeq \Omega_0[(1-\alpha) - \alpha \cos{(2\pi x/l_x)} \cos{(2\pi y/l_y)}]$, with $\alpha=\delta/\Delta_0\sim 0.2$ for realistic experimental parameters, see Fig. \ref{Fig1}(c) \cite{Stellmer13, Chen19}. We therefore name it ``modulation laser''. Band structures analogous to the one depicted in Fig. \ref{Fig1}(b) are obtained for large values of $\alpha\Omega_0/h\gtrsim 20 t=10.7$ kHz and spatial periodicities of the Raman coupling of several lattice sites \cite{footnote2}. The necessary patterns can be projected by combining a spatial light modulator and an optical system of moderate optical resolution, ensuring a large flexibility.\\

\emph{Magic configurations.} The emerging band structures are sensitive to the spatial modulation and the strength of the laser coupling. 
Typically, a system with weak Raman coupling ($\Omega_0 \alpha/h \lesssim t$)  hosts a large number of extended hybridized bands.  Enhanced coupling strength ($\Omega_0 \alpha /h \approx 10 t$) tends to foster band narrowing. 
Remarkably,  there exist  magic configurations of  our considered bilayer square lattice for which special band structures emerge -- quasi-flat bands surrounded by dispersive   Dirac cone spectra with controllable Dirac velocities. We quantify flatness $F$ of a band by the ratio between its width and the dispersion of neighboring bands (cf. \cite{Tarnopolsky19}). This ratio and the  emergence of quasiflat bands can be understood and calculated in perturbation theory (see SM).

The configuration $\Theta(4,4)$ corresponds to  the smallest  bilayer supercell, consisting of $(2 \times 8)$ sites, supporting this  band structure.   Fig.~1(b)  shows  the resulting spin degenerate bands around $E/t=-\Omega_0(1-\alpha)$ for an exemplary case with strong Raman coupling  $\Omega_0 \alpha/h=20 t$, $\alpha=0.2$ and vanishing flux $\gamma=0$. 
Increasing the flux to the simplest experimentally attainable value of $\gamma=0.8$ shares many of the features of the fluxless case. 
Narrow groups of bands, that are well separated from each other, are formed for sufficiently large $\Omega_0 \alpha$ at the energies $\pm\Omega_0$, $\pm\Omega_0(1-\alpha)$ and $\pm\Omega_0 (1-2\alpha)$ (see Fig.~S3 in \cite{eff-ni}). The full spectrum of the system is symmetric around $E=0$, so we only discuss below the band structure for  $E<0$.  
A six band manifold, close to the energy $-\Omega_0(1-\alpha)$ and well separated from nearest neighboring bands by energy $\Omega_0 \alpha$, is shown  along the high-symmetry points in Fig.~2 for (a) $\Omega_0 \alpha/h=2t$ and (b) $\Omega_0 \alpha/h=20t$. Within this six band manifold, the two middle bands closest to $-\Omega_0(1-\alpha)$  become quasi-flat upon increasing $\Omega_0 \alpha$, and are surrounded by a pair of dispersive Dirac cone spectra. 
 The identification of quasi-flatness follows from noticing that the dispersion of the middle bands (which is concentrated towards the edges of the Brillouin zone) is very small $\sim 0.2t$ compared with the bandwidth of its immediately neighbouring bands ($\sim 4t$) for $\Omega_0 \alpha/h=20t$. The flatness $F$ of these bands, calculated in perturbation theory, is proportional to $t/(4 \Omega_0 \alpha)$ and equal to $\sim 0.02$ in the example above. Such values of $\Omega_0$ and $\alpha$ are feasible in our described experimental setup, leading to tunable bandwidth for the quasi-flat bands.
Figs.~2 also shows the associated density of states (DOS), which is given by $D(E)=L^{-d/2}\sum_i\left(E-E\left({\bf k}_i\right)\right)$. 

Interestingly, band structures similar to the $\Theta(4,4)$ case appear when $l_x=l_y=4\nu d$, with $\nu$ integer. This can be explained by treating the intra-layer tunneling as a perturbation to the inter-layer tunneling.
As explained in detail in the Supplemental Material \cite{eff-ni}, the nodal lines of the periodic modulation determine a bilayer Lieb lattice of sites.
The two layers are  energetically well separated  with on-site energies $\pm\Omega_0(1-\alpha)$, respectively.
The perturbation then induces tunnelings within the Lieb lattice topology, which at first order are composed of nearest neighbor tunneling matrix elements within a single layer. The Lieb lattice in its simplest form \cite{Lieb89} is known to host a pair of Dirac cones intersecting at a single $\bf{k}$ point on a completely flat band, the Dirac point. The dispersion of the flat bands in the full model described by Eq.~\eqref{eq:H} originates from higher-order contributions in perturbation theory.
More generally, for $l_{x(y)}=4 \nu_{x(y)}$, where $\nu_{x(y)}$ are positive integers, a similar argument shows that the system can be effectively described by  super-Lieb lattices with a supercell of  $2(\nu_x+\nu_y)-1$ sites.
Changing the periodicity of the Raman coupling in general leads to band structures without the above combination of  Dirac spectra and flat bands \cite{eff-ni}.

\begin{figure}[t!]
\centering
\includegraphics[clip,width=1.0\columnwidth]{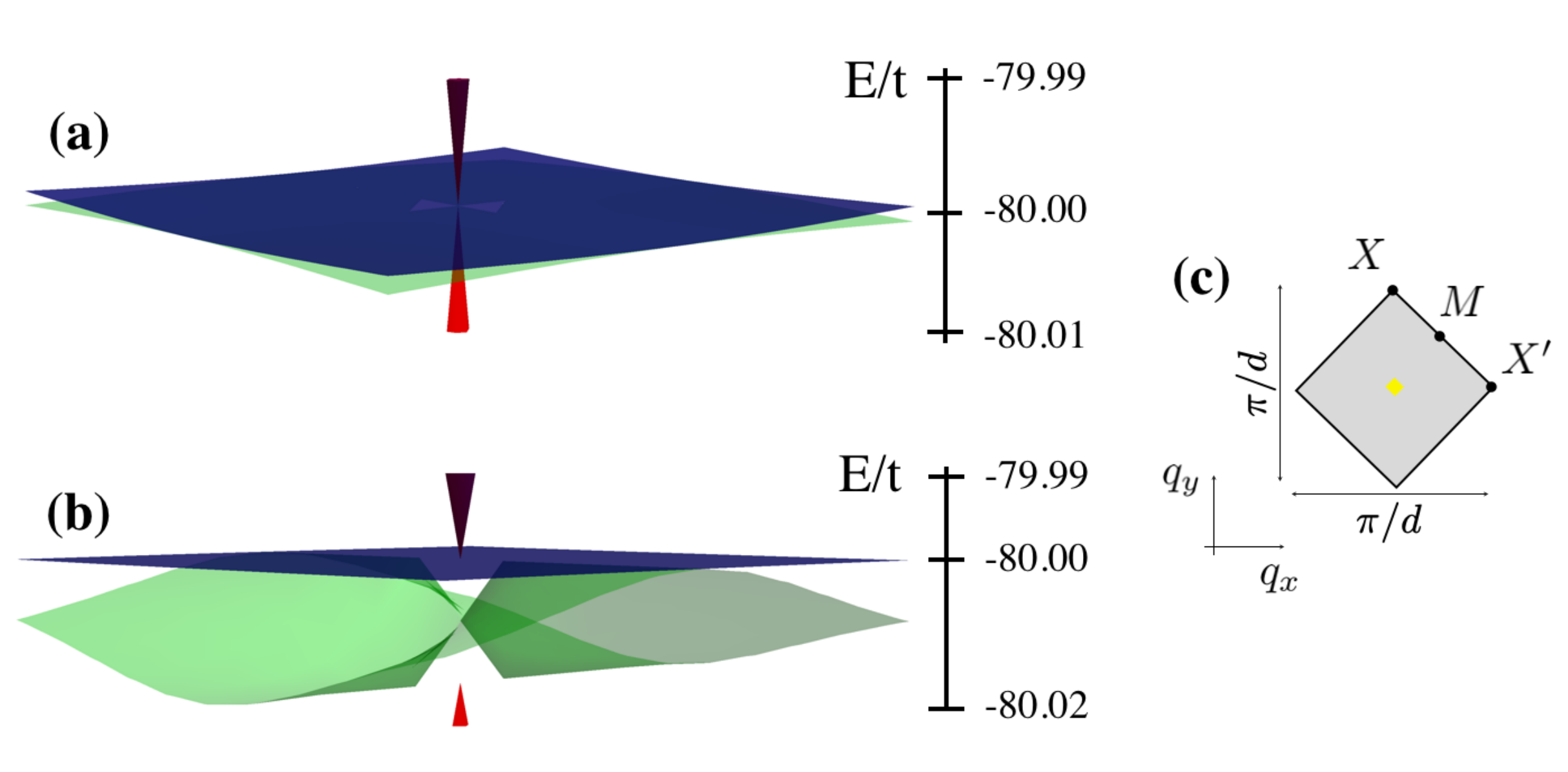}
\caption{Gap opening and  Dirac cone widening due to the artificial flux $\gamma$. Four bands (Dirac cones (purple and red) and quasi-flat bands (green and blue)) are shown in the vicinity of $E/t=-\Omega_0(1-\alpha)=-80$, for the $\Theta(4,4)$ system. The parameters and color scale of the bands are identical to those of Fig.~\ref{Fig2}(b), except for the magnetic flux that corresponds to (a) $\gamma=0$ and (b) $\gamma=0.8$. The energy surfaces are rotated for visibility, and only a small region centered at the $\Gamma=(0,0)$ point in the Brillouin zone (yellow area  in (c), corresponding to $\sim 4\%$ of the complete Brillouin zone) is depicted.}
\label{FigS3a}
\end{figure}

An additional control parameter in our system is the artificial magnetic flux $\gamma$. It affects the band structure in a number of ways. As earlier, we focus on the case $\Theta_M=\Theta(4,4)$ and the $6$ bands closest to $E/t=-\Omega_0(1-\alpha)$. Increasing $\gamma$ leads to strong band narrowing. More interestingly, a non-zero $\gamma$ opens a local gap between the two quasi-flat bands at the $\Gamma=(0,0)$ point in the Brillouin zone. Moreover, the lower Dirac cone detaches from the lower quasi-flat band. This is reminiscent of the effect of a staggered chemical potential in the Lieb lattice~\cite{Shen10}. The upper quasi-flat band remains pinned exactly at the  central energy $\left(E/t=-\Omega_0\left(1-\alpha\right)\right)$  around the $\Gamma$ point, and the upper Dirac cone remains gapless. 
Typical band configurations for two values of $\gamma$ are shown in Fig.~\ref{FigS3a}.  Interestingly, the flux $\gamma$ also controls the Dirac velocity of the cone which decreases with increasing flux. This result can be easily obtained from the perturbative mapping to the Lieb lattice \cite{eff-ni} where, for small to intermediate values of $\gamma$, the dominant nearest neighbour tunneling (proportional to the Dirac velocity) is $\sim \cos(\gamma/2)$. The Dirac velocity reduces to zero in the limit $\gamma \rightarrow \pi$ and the Dirac band becomes very narrow. Hence, both the quasi-flat bands in between the Dirac spectra, as well as the Dirac spectra themselves can be controlled in our scheme.

{\it Conclusions and Outlook.} The basic element in the physics of TwBLG is the creation of large unit cells by rotating two layers with respect to each other. Around the magic angles, small rotations have a dramatic effect on the band structure of these systems. In this Letter, we have discussed a versatile method to create a new class of systems with controllable supercell structures for  cold Fermi  gases trapped in optical lattices. The size of the supercells is easily tunable and should allow addressing whether the physics of TwBLG is uniquely related to their macroscopic periodicity or indeed can be accessed for small unit cells. An inherent advantage of our optical-lattice-based construction is the possibility to modify over a wide range the interlayer coupling, which is controlled by a combination of optical Raman transitions and excited-state light shifts. We have shown that a  square lattice synthetic bilayer displays a band structure that can be easily engineered by modifying the spatial periodicity, strength and Peierls phase imparted by the Raman lasers.As a result, \emph{magic} periodicites of the inter-layer coupling lead to the emergent quasi-flat band physics.

The existence of identical scattering lengths parameterizing interactions between the atoms in the four internal states allows simulating the effect of both intra- and inter-layer interactions in the synthetic bilayer structure. The interacting Hamiltonian can be written as $H_\mathrm{I}=U/2 \sum_{{\bf r}} n({\bf r}) ( n({\bf r})-1)$ where $n({\bf r})= \sum_{m,\sigma} a_{m \sigma}^{\dagger}({\bf r}) a_{m \sigma} ({\bf r})$, is the occupation of site $({\bf r})$ of the square optical lattice. The magnitude of $U$ could be tuned by varying the transverse confinement. In particular, choosing a value of $U$  smaller than $\Omega_0 \alpha$ but much larger than the bandwidth of the quasi-flat band should allow achieving the strongly interacting regime in the latter. Projection of interactions onto the quasi-flat and  hybridizing bands leads to extended Hubbard models with large on-site interactions as well as other terms, such as correlated tunneling. In this respect, it could be advantageous to open a hard gap between the quasi-flat bands and neighboring dispersive Dirac bands. This can be done \emph{via} additional mechanisms, such as lattice dimerization or spin-orbit coupling \cite{Tymek20}. Probing such interacting systems at partial filling could potentially shed new light into theoretical debates on strongly correlated phenomena in twisted materials, such as unconventional superconductivity \cite{Isobe18,You18,Lin19,Gonzalez19} and topological order \cite{Park19,Song19,Ma19}. Finally, extending our approach to other lattice structures represents an exciting perspective for future studies.

\begin{acknowledgements}
We acknowledge funding from European Union (ERC AdG NOQIA-833801), Ministerio de Ciencia, Innovaci\'{o}n y Universidades (FISICATEAMO FIS2016-79508-P, QIBEQI FIS2016-80773-P, FIS2017-86530-P, QuDROP FIS2017-88334-P, and Severo Ochoa SEV-2015-0522), Deutsche Forschungsgemeinschaft (Research Unit FOR2414, Project No. 277974659), Polish National Science Centre (Symfonia Grant No. 2016/20/W/ST4/00314), Generalitat de Catalunya (SGR1341, SGR1381, SGR1646, SGR1660, and CERCA program), Fundaci\'{o} Privada Cellex, Fundaci\'{o} Mir-Puig, Fundaci\'{o}n Ram\'{o}n Areces, and European Social Fund, EU FEDER, European Union Regional Development Fund - ERDF Operational Program of Catalonia 2014-2020 (Operation Code: IU16-011424). AC acknowledges additional support from the Talent Research program of the Universitat Aut{\`o}noma de Barcelona, TS from the Secretaria d'Universitats i Recerca de la Generalitat de Catalunya and the European Social Fund, RWC   from the Polish National Science Centre (NCN) under  Maestro Grant No. DEC-2019/34/A/ST2/00081,  LT from Ministerio de Ciencia, Innovaci\'{o}n y Universidades (RYC-2015-17890), and IF and DR from the Fundaci{\'o} Cellex through a Cellex-ICFO-MPQ postdoctoral fellowship.
\end{acknowledgements}

\section{Supplemental material}
\beginsupplement

\subsection{Effective models \emph{via} perturbation analysis: emergent generalized Lieb lattices}
The origin of the almost flat (weakly dispersive) bands centered around energies $\pm (1-\alpha)\Omega_0 $ for certain magic periodicities (multiples of 4 lattice spacings) of strong Raman coupling can be understood \emph{via} a perturbative treatment of the Hamiltonian Eq.~(1) in the main text. For this regime $\Omega({\bf r})\gg 1$, it is fruitful to diagonalize the Raman coupling term using the operators $c_{\pm 1/2,\sigma}({\bf r})= \exp[\mp \tfrac i2 {\mathbb \gamma}\cdot {\bf r}](a_{1/2,\sigma}({\bf r})\pm a_{1/2,\sigma}({\bf r}))/\sqrt 2$. Setting the energy scale $t=1$, the Raman interlayer coupling then takes the form of a chemical potential
$H_{\mathrm{inter}}= \mp \sum_{{\bf r},\sigma\pm \frac 12}  \Omega({\bf r})  c_{\pm 1/2,\sigma}^{\dagger}({\bf r})c_{\pm 1/2,\sigma}({\bf r})$, while the intralayer perturbative coupling becomes
\begin{multline}
H_{\mathrm{in}}= - \sum_{{\bf r},\hat \mu=\hat x,\hat y}\sum_{m,m',\sigma\pm \frac 12}  c_{m',\sigma}^{\dagger}({\bf r+\hat \mu})\left(\cos\frac{\gamma}2 \mathbbm 1 \right.\\
+ \left.i \sin\frac{\gamma}2 \sigma^x\right)_{m'm} c_{m,\sigma}({\bf r}) +\mathrm{H.c.}
\label{Eq:S2}
\end{multline}
This describes a coupled bilayer system with intralayer tunneling $-\cos\left(\gamma/2\right)$
and interlayer  tunneling $-i\sin\left(\gamma/2\right)$. Recall that the two layers actually correspond to different internal states on a single physical layer, so the mixing term is an effective  spin-orbit coupling.  The two layers have different on-site chemical potentials. We choose the lower layer to host the $c_{+1/2,\sigma}$ fermions and have potential $-\Omega({\bf r})$.

\begin{figure}[t!]
\centering
\includegraphics[clip,width=0.5\columnwidth]{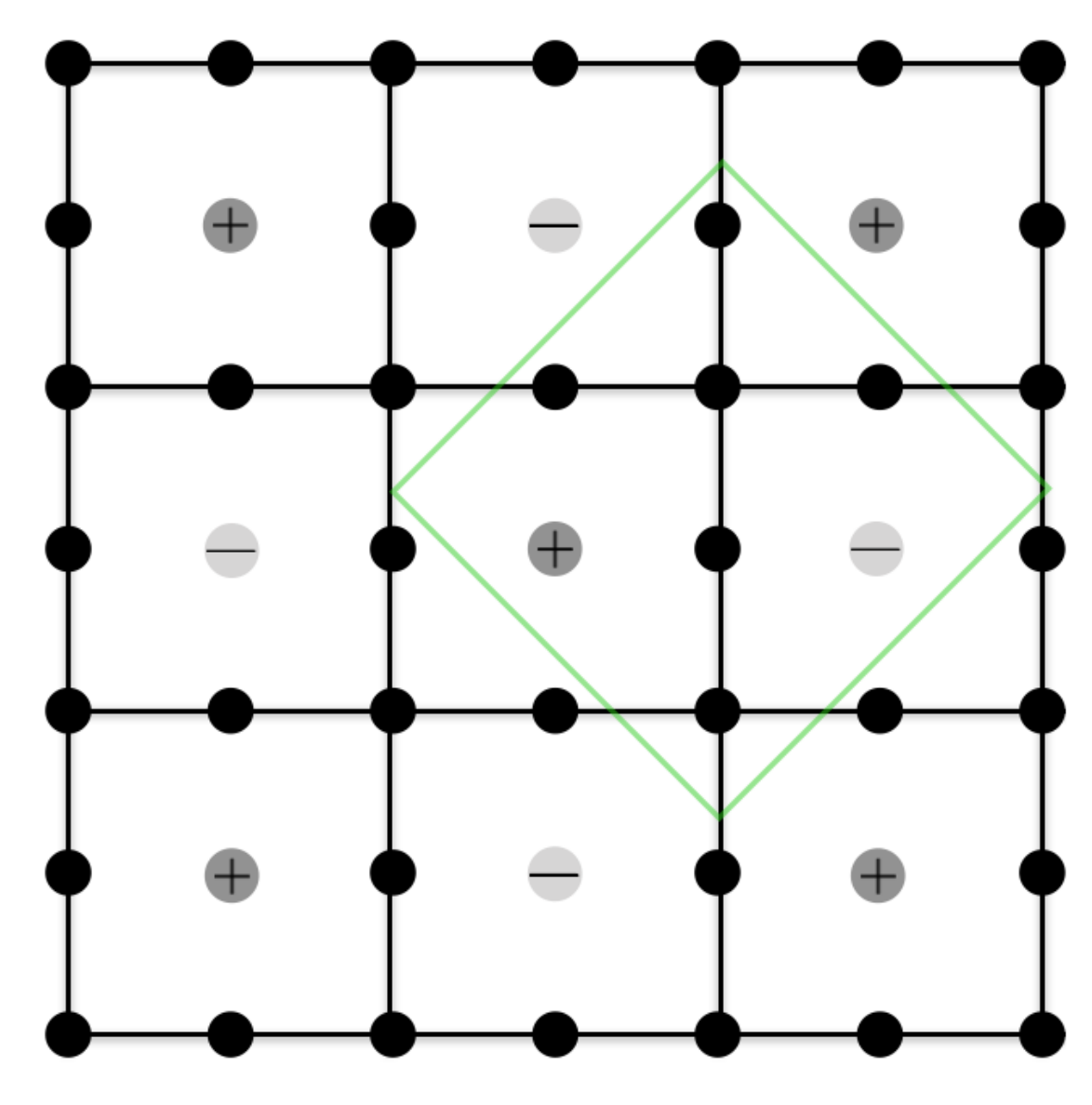}
\caption{ Top view of the lattice for a modulation $\Omega(x,y)\simeq \Omega_0[(1-\alpha) - \alpha \cos{(2\pi x/l_x)} \cos{(2\pi y/l_y)}]$, with $l_x, l_y = 4$. Black sites correspond to $\Omega({\bf r})= \Omega_0(1-\alpha)$, grey sites marked with $\pm$ correspond to sites with $\Omega({\bf r})= \Omega_0(1-\alpha \pm \alpha)$. The latter are energy forbidden sites. Green boundary: unit cell of the lattice. Black lines denote the tunneling between the Lieb lattice sites generated. They are responsible for the main features of the band structures centered at $\pm \Omega_0(1-\alpha)$ described in the main text: a flat band intersected by the Dirac cones. The effects of the grey sites are only taken into account in second order perturbation theory (see Fig.~\ref{FigS2}). }
\label{FigS1}
\end{figure}

The perturbation is highly effective  if the potential $\Omega({\bf r})$ displays equipotential connected regions, i.e. if the nodes of the periodic modulation (cosines) lie on the lattice. This certainly occurs when $l_x$ and $l_y$ are multiples of four. For concreteness, we discuss here the case $l_x=l_y=4$ and $0< \alpha < 1$.
In this case the nodal lines of the periodic modulation determine a Lieb lattice (see Fig.~\ref{FigS1}) of sites with the same on-site energy. Furthermore, we focus on the layer with $m=1/2$  and determine the spectrum centered around $-(1-\alpha) \Omega_0$ \cite{comment1}.  At first order, the perturbation Eq.~(\ref{Eq:S2}) partially lifts the degeneracy by inducing tunneling between sites on the Lieb lattice (black lines in Fig.~\ref{FigS1}).

Nearest neighbor tunneling on a Lieb lattice leads to a three band energy spectrum consisting of a completely flat band containing a Dirac point at which a pair of dispersive bands (with energy respectively higher and lower than the flat band)  intersect.  The Dirac point is located at the corner of the Brillouin zone $(\pm\pi/2,\pm\pi/2)$.
This explains the origin of the band structure around energy $-\Omega_0(1-\alpha)$ (and by analogy around $\Omega_0(1-\alpha)$) as shown in the main text). However, although the Lieb lattice has a unit cell of 3 sites (1 corner and two bridge sites), our full Hamiltonian has a unit cell containing six Lieb lattice sites (one choice is shown in Fig.~\ref{FigS1}).
This  full unit cell is recovered already in second order perturbation theory. The doubling of the Lieb lattice unit cell leads to folding of the Brillouin zone. Therefore, the two Dirac cones~\cite{comment2} in our system are at wave vector ${\bf k}=(0,0)$. Moreover, the spectrum around $-\Omega_0(1-\alpha)$ consists of six bands.

\begin{figure}[t]
\centering
\includegraphics[clip,width=0.7\columnwidth]{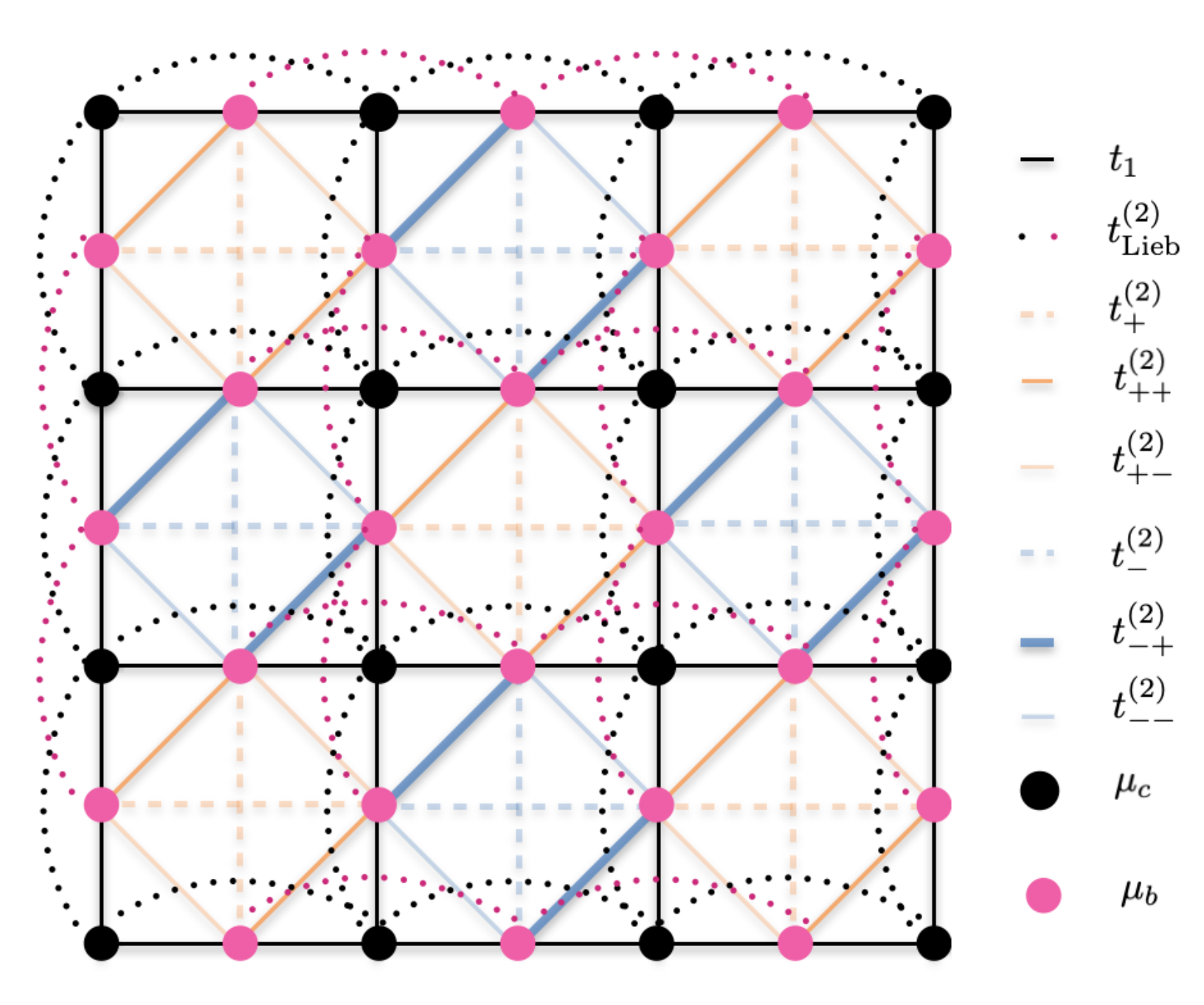}
\caption{Modified effective Lieb lattice for the energy band centered at $-\Omega_0(1-\alpha)$ emerging up to second order of perturbation in the tunneling of Eq.~(\ref{Eq:S2}).  Black full lines are the first order effect. The second order contributions  stem from virtual tunneling to forbidden energy states both within the single layer with $m=1/2$ (grey sites in Fig.~\ref{FigS1}) as well as from spin-orbit coupling interaction coupling to the other $m=-1/2$ lattice (grey sites in Fig.~\ref{FigS1} as well as Lieb lattice sites.  As a result for $\gamma \neq 0,\pi$,  the chemical potential within a layer takes two different values $\mu_c$ on corner (black) and $\mu_b$ on bridge sites (pink). Additionally a periodic pattern of next nearest and third neighbor tunneling is generated as shown by the colored lines (both full and dotted).  The unit cell is that of the original lattice shown in Fig.~\ref{FigS1}. The values of the appropriate tunnelings are written in the text: $t_1$(solid black lines, first order tunneling on the Lieb lattice), $t_{\rm Lieb}^{(2)}$ (black and pink dotted lines), horizontal and vertical tunnelings $t_{\pm}^{(2)}$ between bridge sites over forbidden sites with $\Omega({\bf r}) = \Omega_0(1-\alpha \pm \alpha)$ (dotted orange ($t_{+}^{(2)}$) and blue ($t_{-}^{(2)}$) lines). Finally, tunnelings between next neighboring bridge sites are given by $t_{++}^{(2)}$ (thick orange) and $t_{+-}^{(2)}$ (orange) and $t_{-+}^{(2)}$ (thick blue) and $t_{--}^{(2)}$ (blue).
}
\label{FigS2}
\end{figure}
\begin{figure}[t]
\centering
\includegraphics[clip,width=1.0\columnwidth]{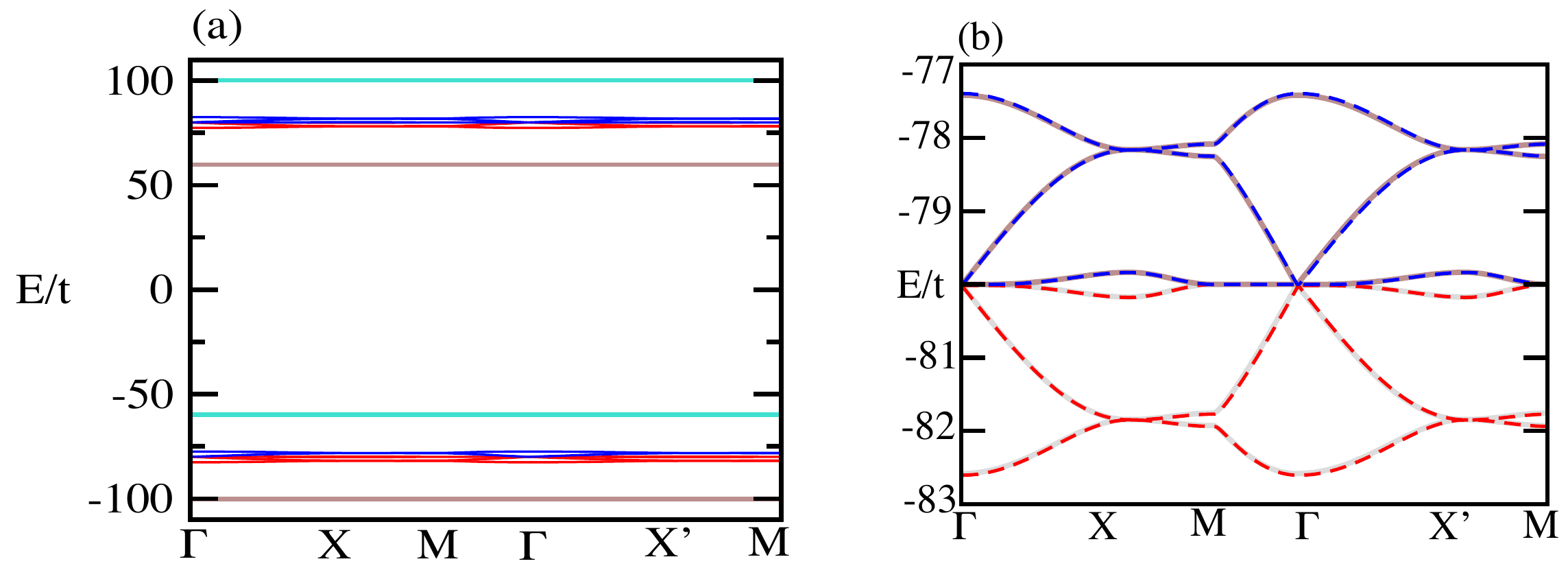}
\caption{(a) Full band structure of $\Theta(4,4)$ for $\Omega_0 \alpha/h=20 t$, with $\alpha=0.2$ and $\gamma=0.8$. (b) Zoom of the energy window around $E/t= -\Omega_0(1-\alpha)=-80$ showing the comparison of exact (solid line) and perturbative (dashed line) results.}
\label{FigS3}
\end{figure}

The detailed band structures, with quasi-flat bands,  shown in the main text in Fig.~2, can be well recovered in second order perturbation theory for large $\Omega_0(1-\alpha)$, as seen in Fig.~\ref{FigS3}. The modification of the band structure is due to an additional periodic pattern of tunnelings in the Lieb lattice of the same periodicity of the supercell $\Theta(4,4)$ generated by virtual tunneling to energy forbidden sites as shown in Fig.~\ref{FigS2}. The various terms described in Fig.~\ref{FigS2} are gathered below.

\begin{gather}
t_{\rm 1} = - \cos \frac{\gamma}{2} , \;\;
t_{\rm Lieb}^{(2)} =\frac{\sin ^2\left(\frac{\gamma }{2}\right)}{2 (1-\alpha ) \Omega _0} \label{Eq:S3}\\
t_{+}^{(2)} = \frac{\sin ^2\left(\frac{\gamma }{2}\right)}{2 (1-\alpha ) \Omega _0+\alpha  \Omega
   _0}+\frac{t_{\rm 1}^2}{\alpha  \Omega _0}\\
t_{++}^{(2)}   = \frac{\sin ^2\left(\frac{\gamma }{2}\right)}{2 (1-\alpha ) \Omega _0+\alpha  \Omega
   _0}+\frac{t_{\rm 1}^2}{\alpha  \Omega _0} + t_{\rm Lieb}^{(2)} \\
t_{+-}^{(2)}   =-\frac{\sin ^2\left(\frac{\gamma }{2}\right)}{2 (1-\alpha ) \Omega _0+\alpha  \Omega
   _0}+\frac{t_{\rm 1}^2}{\alpha  \Omega _0} - t_{\rm Lieb}^{(2)} \\
t_{-}^{(2)} =  \frac{\sin ^2\left(\frac{\gamma }{2}\right)}{2 (1-\alpha ) \Omega _0-\alpha  \Omega
   _0}-\frac{t_{\rm 1}^2}{\alpha  \Omega _0}   \\
t_{-+}^{(2)} =  \frac{\sin ^2\left(\frac{\gamma }{2}\right)}{2 (1-\alpha ) \Omega _0-\alpha  \Omega _0}-\frac{t_{\rm 1}^2}{\alpha  \Omega _0} +  t_{\rm Lieb}^{(2)} \\
t_{--}^{(2)} =  -\frac{\sin ^2\left(\frac{\gamma }{2}\right)}{2 (1-\alpha ) \Omega _0-\alpha  \Omega _0}-\frac{t_{\rm 1}^2}{\alpha  \Omega _0} - t_{\rm Lieb}^{(2)} \\
\mu_c= -\frac{2 \sin ^2\left(\frac{\gamma }{2}\right)}{(1-\alpha ) \Omega _0}
\end{gather}
\begin{gather}
\mu_b = -\left(\frac{1}{2 (1-\alpha ) \Omega _0-\alpha
   \Omega _0}+
   \frac{1}{2 (1-\alpha ) \Omega _0 +\alpha  \Omega _0}\right. \cr
   \left. +\frac{1}{(1-\alpha )
   \Omega _0}\right) \sin ^2\left(\frac{\gamma }{2}\right).  \label{Eq:S11}
\end{gather}

Now, we comment on the evolution of the bands  with the change of the flux $\gamma$.
 From  Eqs.~(\ref{Eq:S3})-(\ref{Eq:S11}) we see that for $\gamma=\pi$ the effective lattice consists of two inequivalent square lattices composed of the bridge and corner sites, respectively, that are completely decoupled from each other. The corner sites form a simple square lattice structure with lattice spacing $2d$, while the bridge sites form a more complicated square lattice. However, neither of these support Dirac cones. These lattices have dispersive  yet very narrow bands since the  tunnel coupling is of the order of $0.01 t$. This is distinct from the opposite limit $\gamma=0$, which is a Lieb lattice with additional tunnelings between bridge sites. Increasing $\gamma$ decreases the total bandwidth of the 6-band system due to the dependence of the dominant coupling $t_1$ Eq.~\eqref{Eq:S3}.  The flat band and Dirac cone subsystem  is modified as $\gamma$ is increased.  In particular,  the upper Dirac cone angle widens as a consequence of the decreasing bandwidth and therefore the Dirac velocity decreases. The upper flat band however  exists  pinned to  the bare chemical potential $-\Omega_0(1-\alpha)$ at the Brillouin zone center for all $\gamma$. We have found that the behaviour of the lower Dirac cone and the opening of a gap to the upper flat band is due to the $C_4$ symmetry breaking modulation in  tunnelling between bridge sites.
Indeed, on one hand this symmetry is explicitly unbroken where there is no gap.
Moreover, we have also checked that setting the other possible factor, i.e. chemical potential staggering to zero does not influence the magnitude of the gap. The unimportance of the staggered chemical potential comes from the fact that the staggering in chemical potential is two orders of magnitude smaller than the staggering in tunneling.

Finally, in the context of twistronics, it is amusing to note that the two disentangled lattices at $\gamma=\pi$ are rotated with respect to each other by $\pi/4$.
 By tuning $\gamma$ towards $0$, their coupling becomes stronger. This leads to a change in the band structure with the formation of Dirac cones that are attracted to each other at the center of the Brillouin zone towards the quasi-flat band. Moreover, it is possible to derive the ratio, $F$, between the bandwidth of the quasi-flat bands within perturbation theory analytically and the all 6 bands of the spectrum as a function of $\Omega_0$, $\alpha$ and $\gamma$. The maximum bandwidth of the quasi-flat bands occurs at the corners of the Brillouin zone and is given by
 \begin{eqnarray}
    \Delta_F = \frac{t^2 \cos ^2\left({\gamma}/{2}\right)}{\Omega_0\alpha} \left(\frac{24\alpha^3-88\alpha^2+106\alpha-32}{3\alpha^3-11\alpha^2+12\alpha-4}\right)
\nonumber\\    
+O(t^3/(\Omega_0\alpha)^2),
 \end{eqnarray}
while the spectrum of all six bands is dominated by the first order term Eq.~(\ref{Eq:S3}) and has the form $\Delta_6=4\sqrt{2} t \cos\left({\gamma}/{2}\right)+O(t^2/(\Omega_0\alpha))$. The relative flatness of the bands can therefore be  calculated as $F=\Delta_F/\Delta_6$.

\begin{figure}[t!]
\centering
\includegraphics[clip,width=0.9\columnwidth]{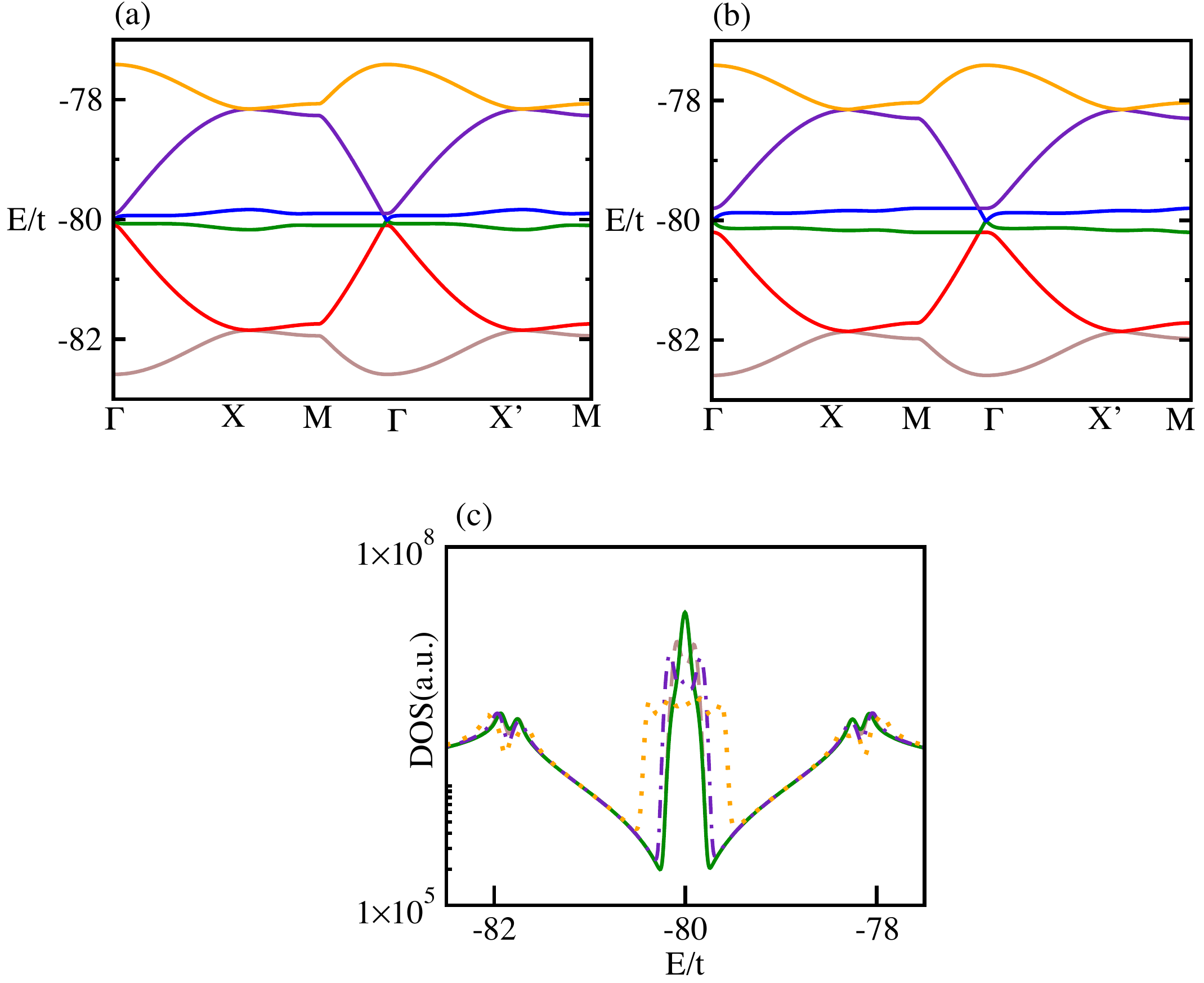}
\caption{Effect of a phase shift in the spatially-dependent periodic Raman coupling. Panels (a-c) corresponds to $\Omega_0\alpha/h$ = $20t$, with $\alpha=0.2$ and $\gamma = 0.8$. Single particle band calculations are shown in the vicinity of $E/t=-(1-\alpha)\Omega_0$ for $\phi_x=\phi_y=0.005$ (a) and 0.01 (b). The DOS (in arbitrary units) for different phases are depicted in (c). The solid, dashed, dashed-dotted and dotted lines are represent the cases for $\phi_x=\phi_y=0$, 0.005, 0.01 and 0.02, respectively. Note the logarithmic scale in (c).}
\label{FigS4}
\end{figure}

\subsection{Effect of phase shifts in the interlayer coupling}
We briefly discuss the effect of allowing a phase in the spatial modulation of the synthetic layer tunneling amplitude, such that  $\Omega(x,y)\simeq \Omega_0[(1-\alpha) - \alpha \cos{(2\pi x/l_x+\phi_x)} \cos{(2\pi y/l_y+\phi_y)}]$. Small phases $\phi_x$ and $\phi_y$ displace the near-degenerate quasi-flat bands at energies $\pm(1-\alpha)\Omega_0$ away from each other. As a result the central peak of the DOS splits into a double peak structure for small values of these phases. Interestingly, a double peak structure in magic angle TwBLG has been reported in previous works \cite{Kerelsky18,Yuan19}. Associated band calculations and DOS are shown in Fig.~\ref{FigS4}.

\subsection{Non-magic configurations}
\begin{figure}[t!]
\centering
\includegraphics[clip,width=0.75\columnwidth]{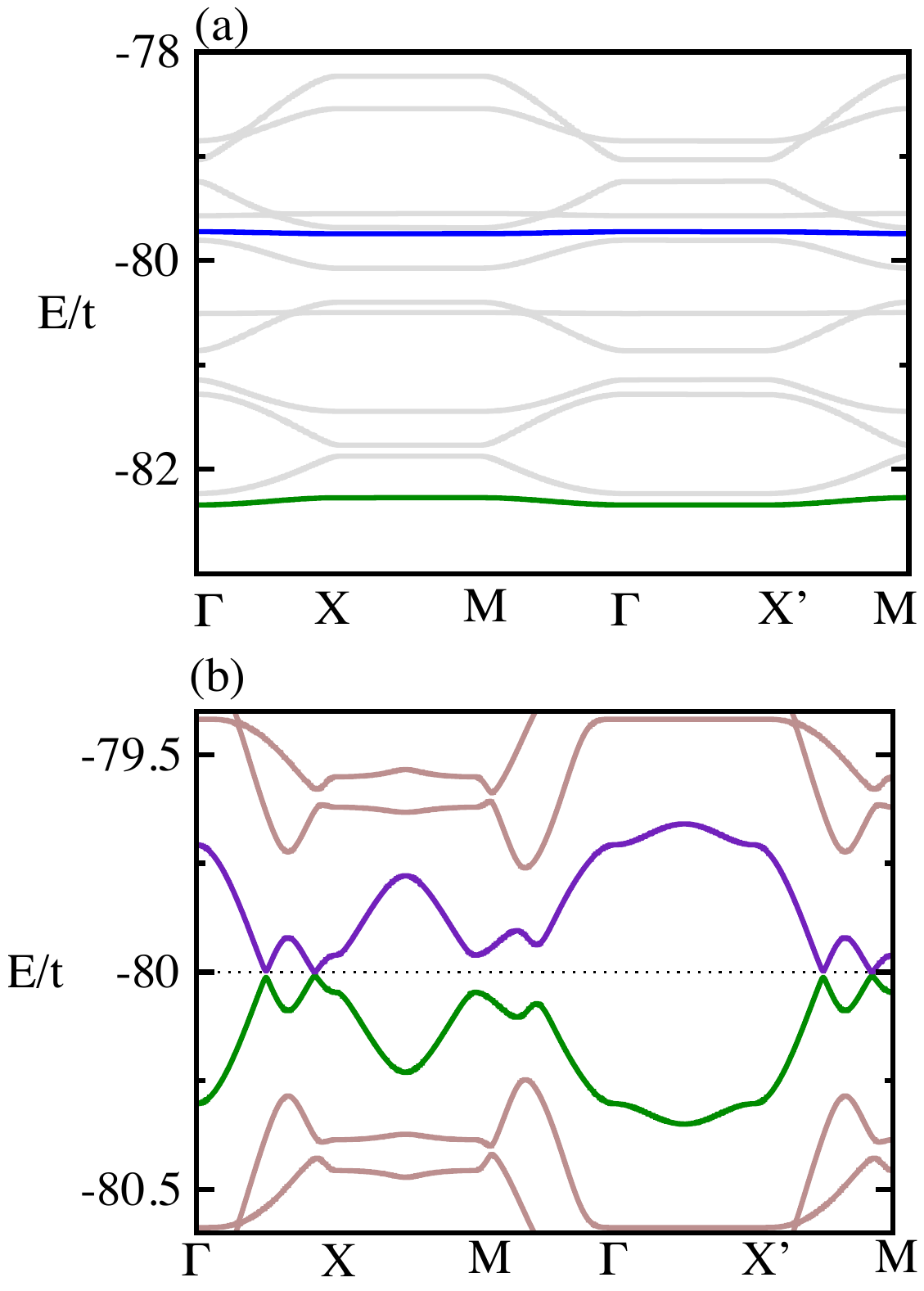}
\caption{Band structures around the energy $-\Omega_0(1-\alpha)$ are depicted for (a) $\Theta(4,7)$ and (b) $\Theta(4,8)$ configuration for $\Omega_0 \alpha/h=20 t$, with $\alpha=0.2$ and $\gamma=0.8$. The dotted line in (b) serves as a guide to the eye.}
\label{FigS5}
\end{figure}

As already mentioned before, the proposed scheme can be exploited to engineer a broad range of band structures by simply manipulating the periodicity of the spatial modulation of the Raman lasers in the square lattice under consideration, or by modifying the lattice structure in itself, along with other parameters, such as Raman coupling strength and magnetic flux. In order to demonstrate that the band structures obtained for $\Theta(4,4)$ are not generic, we illustrate exemplary results for the configurations $\Theta(4,7)$ and $\Theta(4,8)$ in Fig.~\ref{FigS5}(a-b). The chosen parameters have been selected for experimental convenience. We again focus in the vicinity of $E/t=-\Omega_0(1-\alpha)$. While $\Theta(4,7)$ supports a branch of isolated or hybridised flat bands, $\Theta(4,8)$ supports semi-metallic-type bands at $E/t=-\Omega_0(1-\alpha)=-80$.

\begin{thebibliography}{99}
\bibitem{footnote1} To suppress optical transitions to other spin states, an additional laser beam of smaller detuning and producing a large differential light shift can be used \cite{Stellmer11}.
\end{thebibliography}
\begin{thebibliography}{99}
\bibitem{comment1}
Note that the $m=\pm 1/2$ layers are energetically well separated.
\bibitem{comment2} To avoid any confusion, note that in our nomenclature, the Lieb  lattice spectrum (and the spectrum of our model system) has two Dirac cones -- one with "positive" excitation energy and the other with "negative" excitation energy with respect to the quasi-flat band. In contrast, in literature, a shorthand notation is often used whereby the Lieb lattice is said to have a single Dirac cone. This is to distinguish it from lattices where there are two or more inequivalent Dirac points. The latter notation is also motivated by the fact that one is often interested in the (particle) excitations above the flat band.
\end{thebibliography}
\end{document}